%% file: Self-Interference_Cancellation_with_Phase_Noise_Induced_ICI_Suppression_for_Full-Duplex_Systems.tex
\documentclass[conference]{IEEEtran}

\usepackage{graphicx}
\usepackage{color}
\usepackage{placeins}
\usepackage{float}
\usepackage{tabularx,colortbl}
\usepackage{amsmath}
\usepackage{multirow}
\usepackage{hyperref}
\usepackage{mathtools}

\begin{document}
%
\title{Self-Interference Cancellation with Phase Noise Induced ICI Suppression for Full-Duplex Systems}


\author{\IEEEauthorblockN{Elsayed~Ahmed and Ahmed~M.~Eltawil\\}
\IEEEauthorblockA{Electrical Engineering and Computer Science\\
University of California, Irvine, CA, USA\\
\{ahmede,aeltawil\}@uci.edu \\}
\and
\IEEEauthorblockN{Ashutosh~Sabharwal\\}
\IEEEauthorblockA{Electrical and Computer Engineering\\
Rice University, Houston, TX, USA\\
ashu@rice.edu}}

\maketitle

\input{abstract}

%
\IEEEpeerreviewmaketitle

\input{SectionI}

\input{SectionII}

\input{SectionIII}

\input{SectionIV}
\input{SectionV}

\input{References}

\end{document}

%% file: abstract.tex
\begin{abstract}
One of the main bottlenecks in practical full-duplex systems is the oscillator phase noise, which bounds the possible cancellable self-interference power. In this paper, a digital-domain self-interference cancellation scheme for full-duplex orthogonal frequency division multiplexing systems is proposed. The proposed scheme increases the amount of cancellable self-interference power by suppressing the effect of both transmitter and receiver oscillator phase noise. The proposed scheme consists of two main phases, an estimation phase and a cancellation phase. In the estimation phase, the minimum mean square error estimator is used to jointly estimate the transmitter and receiver phase noise associated with the incoming self-interference signal. In the cancellation phase, the estimated phase noise is used to suppress the intercarrier interference caused by the phase noise associated with the incoming self-interference signal. The performance of the proposed scheme is numerically investigated under different operating conditions. It is demonstrated that the proposed scheme could achieve up to 9dB more self-interference cancellation than the existing digital-domain cancellation schemes that ignore the intercarrier interference suppression.
\end{abstract}


%% file: SectionI.tex
\section{Introduction}
Currently, most deployed communication systems operate in half-duplex modes, where bidirectional communication is achieved through either time-division or frequency-division approaches. However, significant improvement in spectral efficiency could be achieved by operating in a full-duplex mode, where bidirectional communications is carried out over the same temporal and spectral resources.

The main limitation impacting full-duplex transmission is managing the strong self-interference signal imposed by the transmit antenna on the receive antenna within the same transceiver. Several recent publications~\cite{Ref1}-\cite{Ref99} have considered the problem of self-interference cancellation in full-duplex systems, showing that the key challenge in practical full-duplex systems is un-cancelled self-interference power caused by a combination of system imperfections. More specifically, the analysis in~\cite{Ref6} demonstrated that in practical full-duplex systems, the transmitter and receiver oscillator phase noise is one of the main bottlenecks that limits the amount of cancellable self-interference. This conclusion is numerically established in~\cite{Ref5} by showing that the channel capacity gain of a full-duplex system significantly decreases as the phase noise becomes stronger. Accordingly, phase noise reduction in full-duplex systems is considered one of the important issues in full-duplex wireless transmission .

In this paper, we consider the problem of self-interference cancellation in full-duplex orthogonal frequency division multiplexing (OFDM) systems in the presence of both transmitter and receiver oscillator phase noise. A digital-domain self-interference cancellation scheme that accounts for the oscillator phase noise is proposed. The proposed scheme increases the amount of cancellable self-interference power by jointly compensating for the transmitter and receiver phase noise associated with the received self-interference signal. 

Generally, the presence of phase noise in an OFDM system introduces intercarrier interference (ICI) at the subcarrier level of the received signal~\cite{Ref7,Ref8}. The strength of the ICI is a function of the received signal strength and the phase noise variance. In full-duplex OFDM systems, due to the strong self-interference power, the ICI associated with the self-interference signal is considered one of the main performance limitations, especially in high phase noise scenarios. Existing digital-domain self-interference cancellation schemes~\cite{Ref1,Ref5,Ref69} ignore the ICI effect, which is reasonable only in low phase noise scenarios. In this paper, the proposed scheme maximizes the amount of canceled self-interference by suppressing the ICI associated with the received self-interference signal.

The problem of ICI suppression in half-duplex OFDM systems has been widely investigated~\cite{Ref7}-\cite{Ref9}. However, the problem of ICI suppression due to phase noise in full-duplex systems is different from the conventional problem in half-duplex systems for two reasons; first, in full-duplex systems, while suppressing the ICI associated with the self-interference signal, the signal-of-interest has to be considered as unknown signal. Second, in full-duplex systems, the self-interference signal is known at the receiver side, thus eliminating the need to use decision feedback techniques to obtain the transmitted signal.

The performance of the proposed scheme is numerically investigated in typical operating environments. The results show that the proposed scheme achieves up to 9dB additional self-interference cancellation compared to the existing digital-domain cancellation schemes.

The remainder of this paper is organized as follows. In Section II, the signal model is presented. The proposed scheme is introduced in Section III. Simulation results and discussions are presented in Section IV. Finally, section V presents the conclusion.

\emph{Notation}: We use $(*)$ to denote convolution, $(.)^H$ to denote conjugate transpose, $E\{.\}$ to denote expectation. We use boldface letters $(\textbf A)$ for matrices, $\textbf A(m,n)$ to denote the element on the $m^{th}$ row and $n^{th}$ column of the matrix $\textbf A$, and $diag(\textbf A)$ to denote a diagonal matrix whose diagonal is constructed from the vector $\textbf A$.

%% file: SectionII.tex
\section{Signal Model}
\begin{figure}[t]
\begin{center}
\noindent
  \includegraphics[width=0.95\columnwidth, height=1.5in]{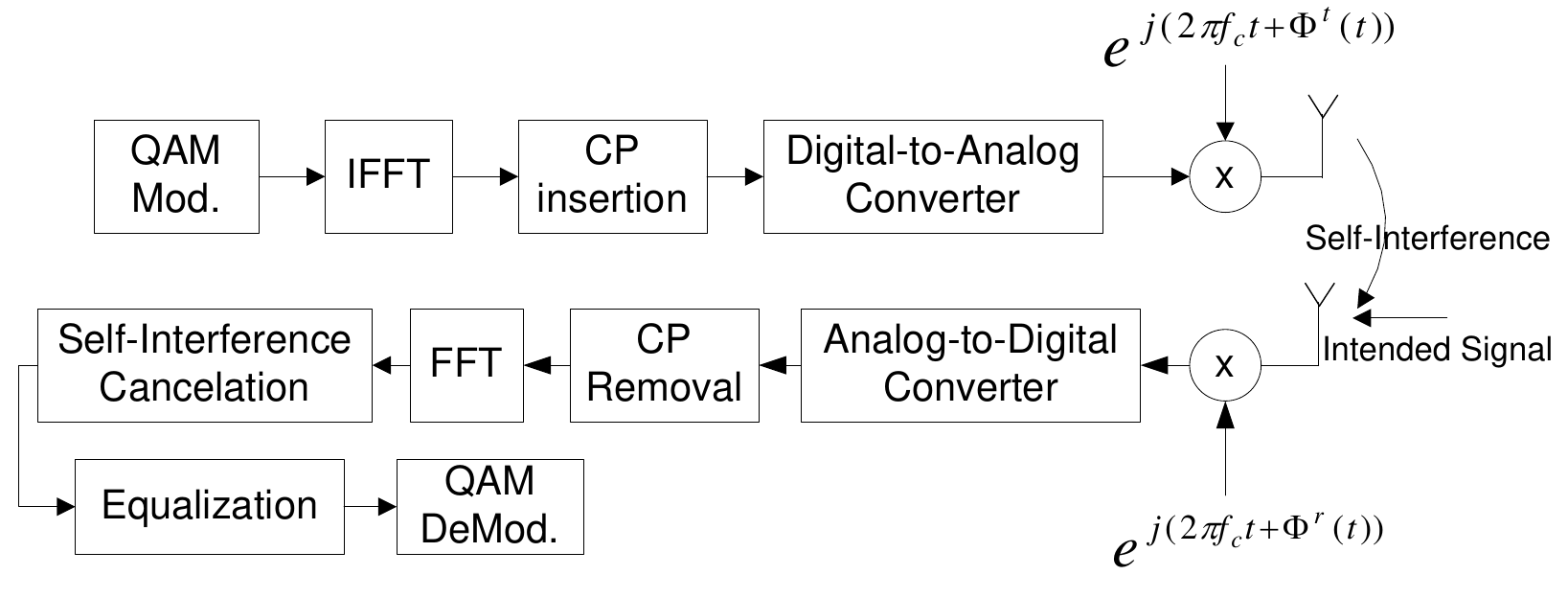}
  \caption{Block diagram of full-duplex OFDM transceiver.\label{Fig1Label}}
\end{center}
\end{figure}

Figure~\ref{Fig1Label} illustrates a block diagram for a full-duplex OFDM transceiver, where the transmitter and the receiver are operating simultaneously over the same carrier frequency. At the transmitter side, the base-band signal is modulated using an OFDM modulator and then up-converted to the carrier frequency $f_c$. The oscillator at the transmitter side is assumed to have a random phase error represented by $\phi^t(t)$. At the receiver side, the received signal consists of the self-interference (the signal from the transmitter in the same transceiver) and the signal-of-interest (the signal to be decoded) is down-converted from the carrier frequency to the base-band. The down-conversion mixer is assumed to have a random phase error represented by $\phi^r(t)$. The base-band signal is then converted to the frequency domain using Fourier transform. In the frequency domain, the self-interference signal is estimated and subtracted from the received signal. Finally, the output of the self-interference cancellation block is equalized and demodulated to restore the transmitted data.

The received base-band time domain signal can be written as
\begin{equation}\label{eq:1}
y_n = \left[\left(x_n^I e^{j\phi_n^{t,I}} * h_n^I \right) + \left(x_n^S e^{j \phi_n^{t,S}} * h_n^S \right)\right] e^{j \phi_n^r} + z_n\text{,}
\end{equation}
where $n$ is the sample index, $x^I$, $x^S$ are the transmitted self-interference and signal-of-interest respectively, $\phi^{t,I}$, $\phi^{t,S}$ are the self-interference and signal-of-interest transmitter phase noise processes, $\phi^r$ is the receiver phase noise process, $h^I$, $h^S$ are the self-interference and signal-of-interest channels, and $z$ is the receiver AWGN noise. The phase noise is modeled as a Wiener process~\cite{Ref10,Ref11} where the phase noise at the $n^{th}$ sample is related to the previous one as $\phi_n = \phi_{n-1} + \alpha$, where $\alpha$ is a Gaussian random variable with zero mean and variance $\sigma^2=4\pi^2f_c^2CT_s$. In this notation $T_s$ describes the sample interval and $C$ is an oscillator dependent parameter that determines its quality. The oscillator parameter $C$ is related to the 3dB bandwidth $f_{3dB}$ of the phase noise Lorentzian spectrum by $C=\Delta f_{3dB}/\pi f_c^2$.

Performing discrete Fourier transform (DFT) on both sides of~\eqref{eq:1} we get
\begin{eqnarray}\label{eq:2}
Y_k &=& \underbrace{\sum_{m=0}^{N-1} \sum_{l=0}^{N-1} X_l^I H_m^I J_{m-l}^{t,I} J_{k-m}^r}_{Y_k^I} \nonumber \\ 
& & + \underbrace{\sum_{m=0}^{N-1} \sum_{l=0}^{N-1} X_l^S H_m^S J_{m-l}^{t,S} J_{k-m}^r}_{Y_k^S} + Z_k \nonumber \\
&=& Y_k^I+Y_k^S+Z_k\text{,}
\end{eqnarray}
where $k$ is the subcarrier index, $N$ is the total number of subcarriers per OFDM symbol, $Y_k^I$, $Y_k^S$ represents the self-interference and signal-of-interest parts of the received signal, $Z_k$ is the Fourier transform of the AWGN noise, and $J^i$, $i\in{[(t,I),(t,S),r]}$ represents the DFT coefficients of the phase noise signal calculated as
\begin{equation}\label{eq:3}
J_k^i = \sum_{n=0}^{N-1} e^{j\phi_n^i} e^{-j2\pi nk/N}\text{.}
\end{equation}
In experimental results published in~\cite{Ref12, Ref13}, it was shown that for full-duplex systems with a strong self-interference line-of-sight component, the self-interference channel follows a Rician distribution with a very large Rician factor (e.g. 25dB to 35dB), and thus can be considered as a frequency-flat channel over wide frequency bands. In the analysis portion of this paper, for simplicity, we assume a frequency-flat channel while developing the cancellation scheme. This assumption is evaluated in the numerical section of the paper and the results show that even at small self-interference channel Rician factor, the performance of the proposed cancellation scheme does not degrade significantly.

Based on the assumption that $H^I$ is is a frequency-flat channel, Equation~\eqref{eq:2} can be simplified as
\begin{eqnarray}\label{eq:4}
Y_k &=& \sum_{l=0}^{N-1}X_l^I H_l^I \sum_{m=0}^{N-1}J_{m-l}^{t,I} J_{k-m}^r + Y_k^S+Z_k \nonumber \\
&=& \sum_{l=0}^{N-1}X_l^I H_l^I J_{k-l}^c + Y_k^S+ Z_k\text{,}         
\end{eqnarray}
where $J^c$ is the DFT coefficients of the combined transmitter and receiver phase noise calculated as the circular convolution of $J^{t,I}$ and $J^r$. From a statistical perspective, $J^c$ represents the DFT coefficients of a phase noise process with an effective 3dB bandwidth $\Delta f_{eff,3dB}$ equal to the sum of both transmitter and receiver phase noise 3dB bandwidths, i.e. $\Delta f_{eff,3dB}=\Delta f_{t,3dB} + \Delta f_{r,3dB}$ or $C_{eff}=C_t+C_r$, where $C_{eff}$ is the effective oscillator parameter.

Rewriting~\eqref{eq:4} in a more detailed form we get
\begin{equation}\label{eq:5}
Y_k = X_k^I H_k^I \underbrace{J_0^c}_{CPE} + \underbrace{\sum_{l=0,l\neq k}^{N-1}X_l^I H_l^I J_{k-l}^c}_{ICI} + Y_k^S+Z_k\text{,}
\end{equation}
where $J_0^c$ is the DC coefficient that acts on all subcarriers as a common phase error (CPE), and the second term represents the ICI associated with the self-interference signal.

%% file: SectionIII.tex
\section{Self-interference cancellation with ICI suppression}
In this section, the proposed self-interference cancellation scheme is introduced. According to~\eqref{eq:5}, total self-interference cancellation requires both the CPE and the ICI components to be suppressed. Existing self-interference cancellation schemes~\cite{Ref1,Ref5,Ref69} only consider the cancellation of the CPE component and neglects the ICI component, which limit the amount of cancellable self-interference power to the ICI level. In this paper, the proposed scheme increases the amount of cancellable self-interference power by considering the suppression of both the CPE and the ICI components.

Generally, self-interference cancellation requires the knowledge of both transmitted self-interference signal ($X^I$) and self-interference channel ($H^I$). Since it is transmitted from the same transceiver, the transmitted self-interference signal is assumed to be known at the reciver side. An accurate estimation for the self-interference channel ($H^I$) as well as the signal-of-interest channel ($H^S$) could be obtained using orthogonal training sequences sent at the beginning of each transmission frame~\cite{Ref1}. In our derivations below, we assume that the self-interference and signal-of-interest signal channels are perfectly known at the receiver.

The proposed scheme consists of four main steps;
\begin{itemize}
\item
Estimating the DC coefficient ($J_0^c$).
\item
Suppressing the CPE component by subtracting $X_k^I H_k^I J_0^c$ from the received signal.
\item
Estimating the remaining phase noise coefficients ($J_i^c, i\neq 0$).
\item
Suppressing the ICI component by reconstructing the signal $\sum_{l=0,l\neq k}^{N-1}X_l^I H_l^I J_{k-l}^c$ and subtract it from the received signal.
\end{itemize}
For the DC coefficient estimation, the least square (LS) estimator is used as follows
\begin{equation}\label{eq:6}
J_0^c = \frac{1}{N_u} \sum_{k=0,k\in U}^{N_u-1}\frac{Y_k}{X_k^I H_k^I} \text{,}         
\end{equation}
where $U$ is a set contains the pilot positions within the OFDM symbol, and $N_u$ is the number of pilot subcarriers. After estimating the DC coefficient, the CPE component is subtracted from the received signal as follows
\begin{equation}\label{eq:7}
Y_k-X_k^I H_k^I J_0^c=\sum_{l=0,l\neq k}^{N-1}X_l^I H_l^I J_{k-l}^c +Y_k^S+Z_k\text{.}         
\end{equation}

In order to perform ICI suppression, the remaining coefficients of $J^c$ have to be estimated. Based on~\eqref{eq:7}, the problem of estimating $J^c$ is considered as a linear estimation problem, where $J^c$ is a parameter vector distributed by Gaussian noise and the signal-of-interest ($Y^S$). For an estimation order $M$ (where $M$ is the number of coefficients to be estimated), Equation~\eqref{eq:7} can be written in a matrix form as
\begin{eqnarray}\label{eq:8}
\left[ \begin{array}{c} B_{l_1} \\ B_{l_2} \\:\\ B_{l_p} \end{array} \right] &=&
\begin{bmatrix} A_{l_1} & ... &A_{l_1+M} \\ A_{l_2} & ... &A_{l_2+M}\\:&:&: \\ A_{l_p} & ... &A_{l_p+M} \end{bmatrix}
\left[ \begin{array}{c} J_{M/2}^c \\:\\J_1^c\\J_-1^c\\:\\ J_{-M/2}^c \end{array} \right] \nonumber \\
& & + \left[ \begin{array}{c} Y^S_{l_1} \\ Y^S_{l_2} \\:\\ Y^S_{l_p} \end{array} \right] 
+ \left[ \begin{array}{c} \gamma^{ICI}_{l_1} \\ \gamma^{ICI}_{l_2} \\:\\ \gamma^{ICI}_{l_p} \end{array} \right] 
+ \left[ \begin{array}{c} Z_{l_1} \\ Z_{l_2} \\:\\ Z_{l_p} \end{array} \right]\text{,} 
\end{eqnarray}
where $B_k=Y_k-X_k^I H_k^I J_0^c$ , $A_k=X_k^I  H_k^I$, and $\gamma^{ICI}$ is the residual ICI beyond the estimation order $M$. The set $[l_1 \ \ l_2 \ \ ... \ \ l_p]$ has to be of length $\geq M$ in order to solve~\eqref{eq:8} for $M$ unknowns. Summarizing~\eqref{eq:8} in a compact form we get
\begin{equation}\label{eq:9}
\bf B=\bf A \bf J^c + \boldsymbol{\eta} \text{,}         
\end{equation}
where $\boldsymbol{\eta}$ represents the effective noise that combines all of the signal-of-interest, the residual ICI, and the AWGN noise. Using~\eqref{eq:9}, the minimum mean square error (MMSE) estimate of $\bf J^c$ is given by~\cite{Ref14}
\begin{equation}\label{eq:10}
\bf J^c=\bf W \bf B\text{,}         
\end{equation}
\begin{equation}\label{eq:11}
\bf W=\bf R_{JJ} \bf A^H (\bf A \bf R_{JJ} \bf A^H + \bf R_{\eta \eta})^{-1} \text{,}         
\end{equation}
where $\bf R_{JJ}$ represents the correlation matrix of the vector $\bf J^c$, and $\bf R_{\eta \eta}$ represents the correlation matrix of the vector $\boldsymbol{\eta}$.

Using the transmitter and receiver oscillator parameters, the correlation matrix $\bf R_{JJ}$ can be calculated as in~\cite{Ref8}
\begin{equation}\label{eq:12}
\bf{R_{JJ}} (p,q)=\sum_{k=0}^{N-1}\sum_{l=0}^{N-1} e^{-4\pi^2 f_c^2 C_{eff} T_s |k-l|/2} e^{-j2\pi(pk-ql)/N}\text{.}         
\end{equation}
Assuming that the data symbols and the AWGN noise are not correlated, the correlation matrix $\bf R_{\eta \eta}$ can be written as 
\begin{eqnarray}\label{eq:13}
\bf R_{\eta \eta} &=& \mathsf{diag}(E\left\{\left|Y_{l_1}^S \right|^2\right\}+E\left\{\left|\gamma_{l_1}^{ICI}\right|^2\right\}+\sigma_z^2 ...... \nonumber \\
& & E\left\{\left|Y_{l_p}^S \right|^2\right\}+E\left\{\left|\gamma_{l_p}^{ICI}\right|^2\right\}+\sigma_z^2)\text{,}         
\end{eqnarray}
where $\sigma_z^2$ is the AWGN noise variance, $E\left\{\left|\gamma_{l_i}^{ICI}\right|^2\right\}$ is the power of the residual ICI at subcarrier $l_i$ calculated as~\cite{Ref8}
\begin{equation}\label{eq:14}
E\left\{\left|\gamma_{l_i}^{ICI}\right|^2\right\}=\sum_{n=0,n>|M|}^{N-1} \bf{R_{JJ}} (n,n)\text{,}         
\end{equation}
and $E\left\{\left|Y_{l_1}^S\right|^2\right\}$ is the power of the received signal-of-interest at subcarrier $l_i$. For simplicity, $E\left\{\left|Y_{l_1}^S\right|^2\right\}$ can be approximated to the average received signal-of-interest power as follows
\begin{equation}\label{eq:15}
E\left\{\left|Y_{l_i}^S\right|^2\right\}=E\left\{\left|X_{l_i}^S H_{l_i}^S\right|^2\right\} = E\left\{\left|H_{l_i}^S\right|^2\right\}\text{,}         
\end{equation}
where the transmitted signal are assumed to be M-QAM modulated with a unity average power.

Finally, the phase noise vector $\bf J^c$ is constructed by placing the $M$ estimated coefficients in their corresponding positions and placing zero elsewhere. At the end, the ICI component is reconstructed as
\begin{equation}\label{eq:16}
ICI_k = \sum_{l=0,l\neq k}^{N-1} X_l^I H_l^I J_{k-l}^c \text{,}         
\end{equation}
and then subtracted from the received signal.

%% file: SectionIV.tex
\section{Simulation results and discussions}
In this section, the performance of the proposed self-interference cancellation scheme is numerically investigated under different operating conditions. The system is assumed to operate in a full-duplex mode, where the wireless terminals are transmitting and receiving at the same time using the same carrier frequency. The transmitted frame consists of 50 OFDM symbols with 64 subcarriers in each symbol. Each OFDM symbol contains 4 pilot subcarriers used for CPE estimation. The carrier frequency $f_c$ is set to 2.4GHz with a system bandwidth of 20MHz. The indoor TGn channel model D~\cite{Ref15} is used to model the self-interference and signal-of-interest channels. The self-interference and signal-of-interest channel's Rician factors are set to 30dB and 3dB respectively.

\begin{figure}[t]
\begin{center}
\noindent
  \includegraphics[width=0.95\columnwidth ,height=2.8in]{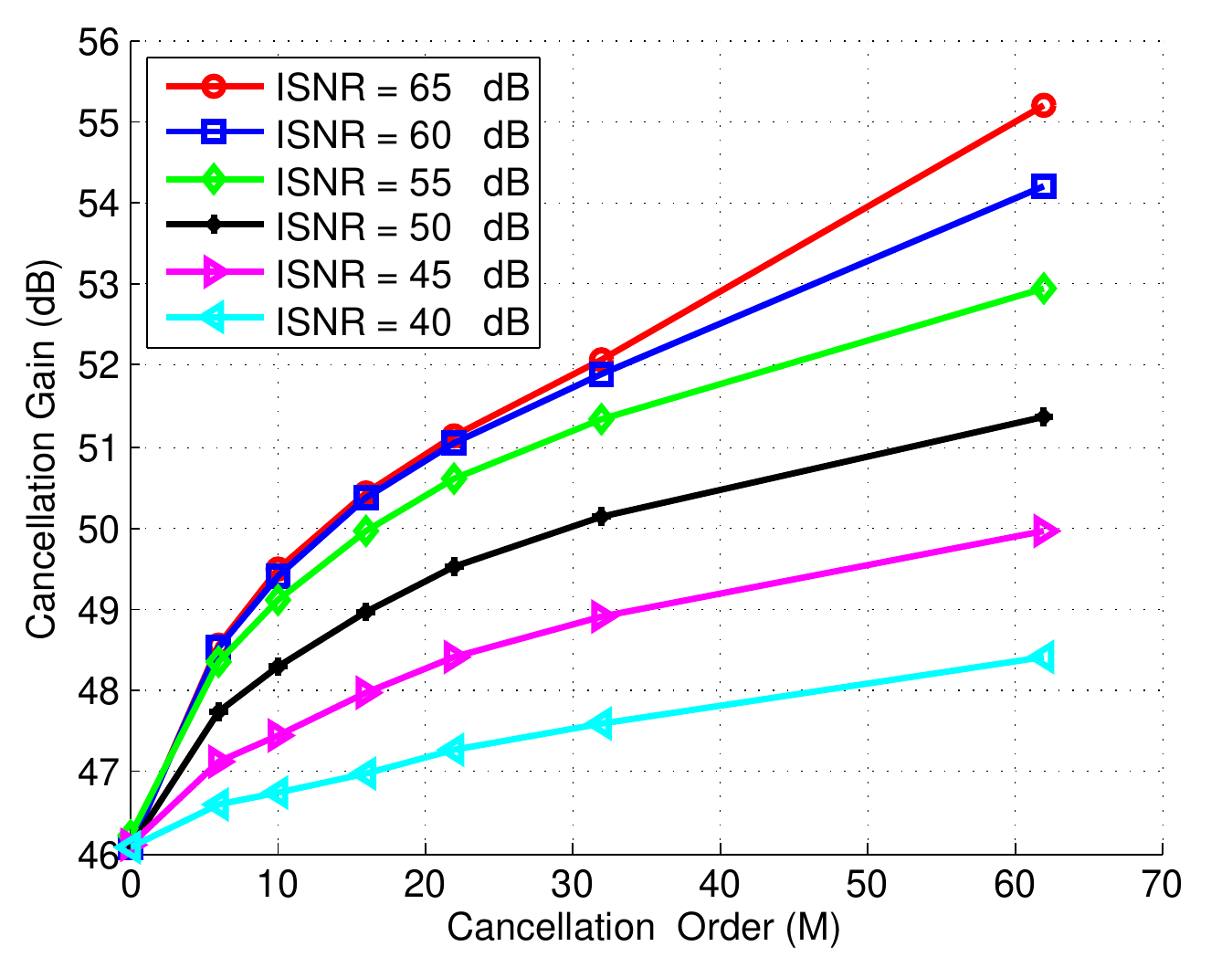}
  \caption{Self-interference cancellation gain at different ISNR values, $C_{eff}=1.64e^{-19}$.\label{Fig2Label}}
\end{center}
\end{figure}

In the first simulation scenario, we investigate the performance of the proposed scheme under different self-interference and signal-of-interest power scenarios. The self-interference \emph{cancellation gain} defined as the remaining self-interference power after cancellation divided by the received self-interference power is used as a performance criterion. In this analysis, the interference to signal plus noise ratio (ISNR) is defined as the power of the received self-interference signal divided by the summation of the signal-of-interest power and the AWGN noise power before self-interference cancellation. Figure~\ref{Fig2Label} shows the self-interference cancellation gain for different cancellation orders ($M$) at different ISNR values. It has to be noticed that; first, $M=0$ means that only CPE component is canceled and there is no ICI suppression (similar to the existing digital-domain cancellation schemes). Second, based on the chosen value of $C_{eff}=1.64e^{-19}$, the total ICI power is around -46dB normalized to the received self-interference power. This is why the cancellation gain at $M=0$ (no ICI suppression) in Figure~\ref{Fig2Label} is around 46dB.

The results show that at high ISNR values, the proposed scheme could achieve up to 9dB more cancellation gain compared to the case where only the CPE component is canceled. The results also show that as the ISNR increases the cancellation gain increases. The reason is that, the mean square error of the MMSE estimator decreases with the decrease of the noise variance (i.e. increase of the ISNR). It has to be noticed that, due to the close proximity of transmit and receive antennas at the same transceiver, full-duplex systems are expected to operate at very high ISNR values, where the proposed scheme achieves its best performance.
\begin{figure}[t]
\begin{center}
\noindent
  \includegraphics[width=0.95\columnwidth ,height=2.8in]{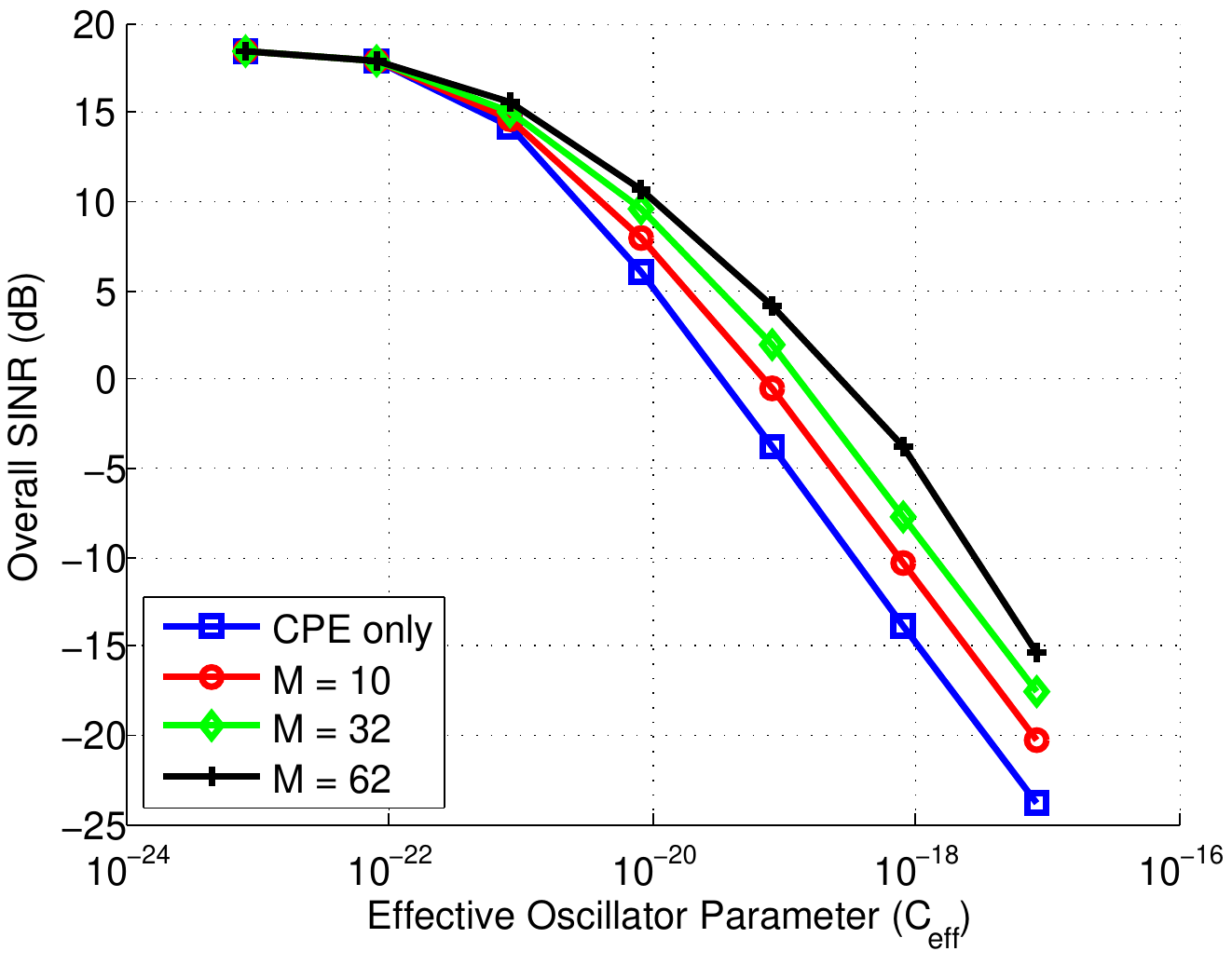}
  \caption{Overall SINR at different phase noise values.\label{Fig3Label}}
\end{center}
\end{figure}

In the following simulation scenario, the overall system performance is investigated by evaluating the overall signal to interference plus noise ratio (SINR). The overall SINR is defined as the signal-of-interest power divided by the summation of the residual self-interference power and the AWGN noise power. Figure~\ref{Fig3Label} shows the overall SINR at different phase noise values (represented by $C_{eff}$). The results show that at high phase noise values, the proposed scheme achieves considerable SINR gain (up to 9dB) compared to the existing cancellation schemes, where only the CPE component is canceled. However, as the phase noise decreases the achieved SINR gain decreases until it reaches 0dB at very low phase noise values. At this point, both the proposed and the existing schemes achieve the same performance. The reason is that, at high phase noise values, the SINR is dominated by the residual self-interference power, which is limited by the ICI component. Therefore, the cancellation gain achieved by suppressing the ICI component will improve the overall SINR. On the other hand, at low phase noise values, the ICI component goes below the AWGN noise level. At this point, only CPE cancellation is sufficient to mitigate the self-interference signal to the AWGN noise level. As a conclusion, the proposed algorithm shows a considerable SINR gain in the region where phase noise dominates system performance, which is the region where current full-duplex systems operate~\cite{Ref6}. 

\begin{figure}[t]
\begin{center}
\noindent
  \includegraphics[width=0.95\columnwidth ,height=2.8in]{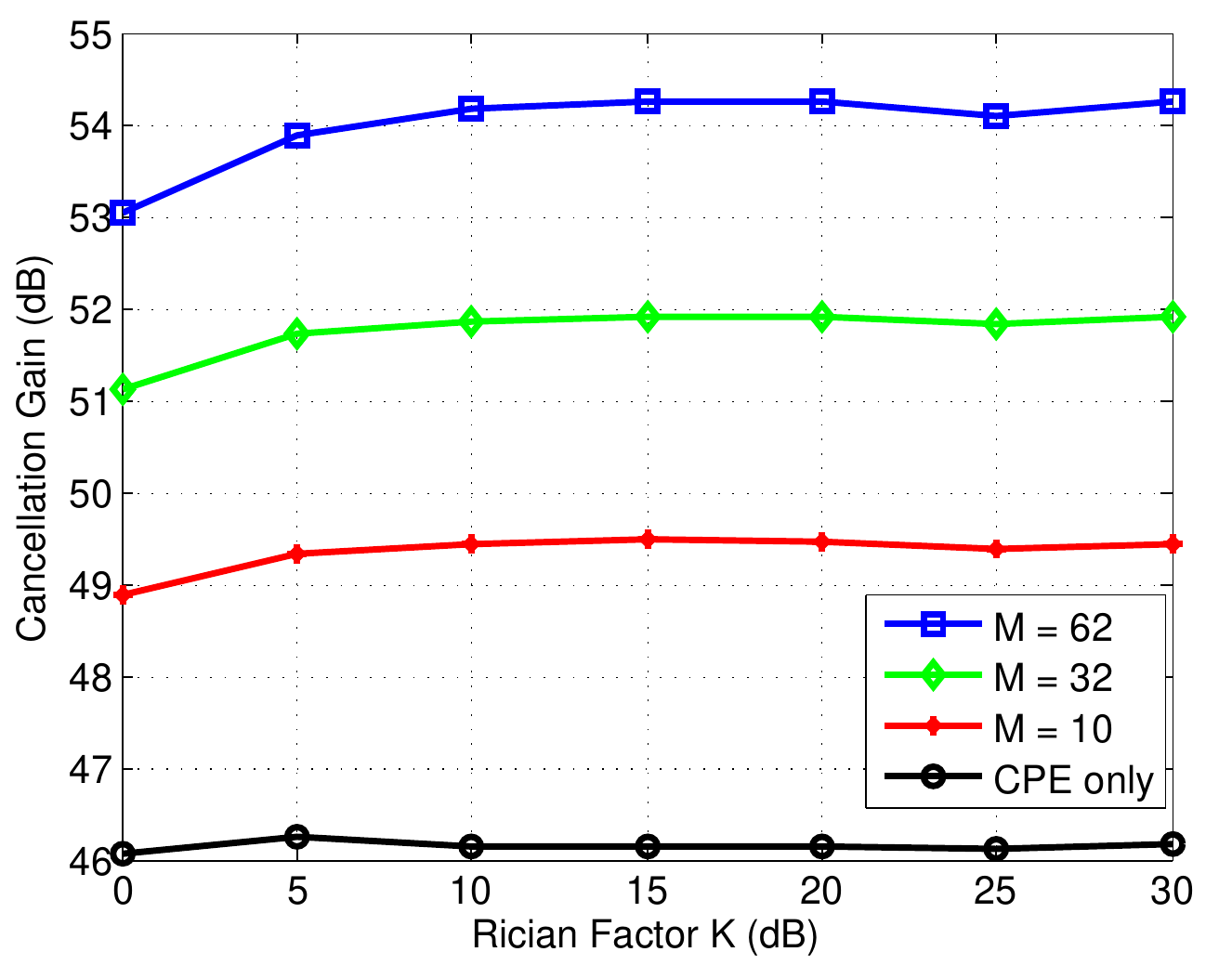}
  \caption{Self-interference cancellation gain at different self-interference channel's Rician factors.\label{Fig4Label}}
\end{center}
\end{figure}

Finally, we investigate the effect of the self-interference channel's Rician factor ($K$) on the performance of the proposed scheme. Figure~\ref{Fig4Label} shows the self-interference cancellation gain achieved by the proposed scheme at different $K$ values. The results show that for high $K$ values, the proposed scheme achieves constant performance. However, for relatively low $K$ values ($<$5dB), the performance is slightly degraded. The reason is that, decreasing $K$ will decrease the validity of the assumption that the self-interference channel is a frequency-flat channel. Despite the performance degradation, the results show that even with $K$ values as low as 0dB, the performance degradation does not exceed 1dB relative to the high $K$ scenarios.

%% file: SectionV.tex
\section{Conclusion}
In this paper, a digital-domain self-interference cancellation scheme for full-duplex OFDM systems is proposed. The proposed scheme increases the amount of cancellable self-interference power by suppressing the effect of both transmitter and receiver oscillator phase noise. The proposed scheme consists of two phases: first, a MMSE estimator is used to jointly estimate the transmitter and receiver oscillator phase noise. Second, the estimated phase noise is used to suppress the ICI caused by the phase noise associated with the incoming self-interference signal. The performance of the proposed scheme is numerically investigated under different operating conditions. The results show that, for phase noise limited full-duplex systems, the proposed scheme could achieve up to 9dB SINR gain compared to the existing schemes that ignore the ICI suppression.